\documentclass[pra,10pt,twocolumn,superscriptaddress,showpacs]{revtex4-1}
\usepackage{amsmath}
\usepackage{latexsym}
\usepackage{amssymb}
\usepackage{mathtools}
\usepackage{braket}
\usepackage{graphicx}
\usepackage{dsfont}
\usepackage[colorlinks=true, citecolor=blue, urlcolor=blue]{hyperref}
\usepackage{float}
\usepackage{amsfonts}

\begin{document}

\title{Two-qubit mixed states and teleportation fidelity: Purity, concurrence, and beyond}

\author{Sumit Nandi}
\email{sumit@iopb.res.in}
\affiliation{Institute of Physics, Sachivalaya Marg,
Bhubaneswar 751005, Odisha, India.}
\affiliation{Homi Bhabha National Institute, Training School Complex, Anushakti Nagar, Mumbai 400085, India.}

\author{Chandan Datta}
\email{chandan@iopb.res.in}
\affiliation{Institute of Physics, Sachivalaya Marg,
Bhubaneswar 751005, Odisha, India.}
\affiliation{Homi Bhabha National Institute, Training School Complex, Anushakti Nagar, Mumbai 400085, India.}

\author{Arpan Das}
\email{arpandas@iopb.res.in}
\affiliation{Institute of Physics, Sachivalaya Marg,
Bhubaneswar 751005, Odisha, India.}
\affiliation{Homi Bhabha National Institute, Training School Complex, Anushakti Nagar, Mumbai 400085, India.}

\author{Pankaj Agrawal}
\email{agrawal@iopb.res.in}
\affiliation{Institute of Physics, Sachivalaya Marg,
Bhubaneswar 751005, Odisha, India.}
\affiliation{Homi Bhabha National Institute, Training School Complex, Anushakti Nagar, Mumbai 400085, India.}

\begin{abstract}
To explore the properties of a two-qubit mixed state, we consider quantum teleportation.
The fidelity of a teleported state depends on the resource state purity and entanglement, as characterized
by concurrence. Concurrence and purity are functions of state parameters. However, it turns out that a state with larger
purity and concurrence, may have comparatively smaller fidelity. By computing  teleportation fidelity, concurrence and purity for two-qubit X-states, we show it explicitly. We further
show that fidelity changes monotonically with respect to functions of parameters - other than
concurrence and purity. A state with smaller concurrence and purity,
but larger value of one of these functions has larger fidelity. These
functions, thus characterize nonlocal classical and/or quantum properties of the state
that are not captured by purity and concurrence alone. In particular, concurrence is
not enough to characterize the entanglement properties of a two-qubit mixed state.
\end{abstract}

%\pacs{03.65.Ud, 03.67.Hk, 03.67.Bg, 03.67.Mn} 

\maketitle

\section{Introduction}
Quantum entanglement has remained a major resource for accomplishing quantum information processing tasks such as
teleportation \cite{fid1}, quantum key distribution \cite{ekert}, secret sharing \cite{hillery} etc. The key idea of teleportation, as  
 proposed by Bennett \textit{et al.} \cite{fid1} in their seminal paper, is to transmit an unknown state to
 a remote location using the entanglement as a resource.
 In the paper, the authors used a singlet state which is known to be maximally-entangled state as a resource. With this resource, an unknown one-qubit state can be transmitted with unit fidelity
 and unit probability - i.e., perfect teleportation.
 In a teleportation scheme, the resource state is used as a channel for transmission to a distant location. 
 The efficiency of a channel is quantified by teleportation fidelity which takes its maximum value for 
  maximally-entangled Bell states.
 A classical channel can also be used to perform teleportation with
 fidelity upper bounded by $\frac{2}{3}$ \cite{Popescu}. A unit fidelity is possible only 
 if the resource state is maximally entangled.\\
In a realistic scenario, a resource state is usually a mixed state. It happens due to the unavoidable noise,
 interaction with the environment which eventually turns a pure state 
into a mixed one. It was Popescu \cite{Popescu} who showed that a mixed state can be sometime useful for teleportation
producing fidelity greater than the classical value.  A mixed state has both classical and quantum properties.
Their interplay can by quite complex. Usually, classicality  of a state is characterized by how pure a state is. Here, by classicality we mean the classical correlations in the state due to the mixing parameters. The popular measures are -  von Neumann entropy, linear entropy, or purity. We shall use the later which can be defined as $\mathcal{P}=Tr[\rho^2]$; it attains its maximum value one for a pure state.
Nonlocal quantum properties are characterized by an entanglement measure, such as concurrence, or negativity.
One may expect that the teleportation fidelity may  depend  not only on the entanglement of a mixed state, but
also classicality. \.{Z}yczkowski \textit{et al.} \cite{Zyczkowski} have shown that states with purity less than $\frac{1}{3}$ are separable. It has also been proved \cite{Bose} that if a state exceeds a certain degree of mixedness as quantified by von Neumann entropy, then it can't be used as teleportation channel. Likewise in \cite{Paulson} the authors have obtained
rank dependent lower bound on concurrence and upper bound on mixedness of a state for the success of the
teleportation protocol. Verstraete \textit{et} \textit{al.} \cite{Verstraete} have found an upper 
bound on fidelity in terms of its entanglement
as measured by concurrence \cite{Wooters} or negativity \cite{Vidal}. As we shall see,
 there exist states with smaller purity or/and concurrence, but higher teleportation fidelity. 
 In this paper, we will obtain analytical expressions
relating three quantities namely, purity, concurrence and fidelity. We go beyond this also. We will analyze the comparative behavior for a class of states, which encompass many well known states. 
%{\color{Blue} A two-qubit mixed state is characterized by 15 parameters, which can be reduced to 9 by applying local unitary transformations. 
%This can be seen in Hilbert-Schmidt basis \cite{Horodecki}, or using spectral decomposition \cite{spectral}.} {\color{Red}Other fact is that a two-qubit mixed state in a joint Hilbert-Schmidt space is characterized by 15 %parameters.}
 The number of independent parameters in this class of states go beyond two quantities -- purity and concurrence -- that are normally used to 
characterize a mixed state. By considering mixed states with fewer parameters, we will show that optimal teleportation fidelity
monotonically increases or decreases with new functions of the state parameters. So these extra functions also characterize
the classical and nonlocal quantum properties of a mixed state. Purity and concurrence are not enough to
fully characterize a state, and to understand the optimum implementation of a communication protocol. For example,
larger purity, or/and larger concurrence does not necessarily imply larger teleportation fidelity. There are other functions of the state parameters that are also needed to charaterize the nonlocal properties of the state as we show less entangled and less pure state can produce better fidelity. 
Although, we shall demonstrate the usefulness of these functions in the context of teleportation fidelity,
same functions play similar role in characterizing the Bell violation \cite{ToBePublished}. Recently there has been similar discussion in the context of entanglement purification \cite{Torres}. It has been shown that the entanglement purification protocol can be enhanced if we have access to the full parameter space of a two-qubit mixed state. Our motivation is slightly different in the sense that we show that nonlocal properties of a mixed state do require physical quantities other than purity and concurrence. 
We present our discussion using X-states as resource states for the protocol. \\~\\
  Recently Mendon\c{c}a \textit{et al.} \cite{fid2} have shown that for 
every two-qubit state there exists a  density matrix parameterized by seven parameters which resembles the alphabet $X$ and popularly known as X-state in literature and  with same purity and concurrence as of the former. Moreover, they have shown that the whole concurrence-purity region of any two-qubit state can be covered with only X-states. Many families of two-qubit states like
Werner state,  Bell states, maximally-entangled mixed state have this kind of structure. X-states were first shown to have interesting properties in a paper by Yu and Eberly \cite{yu}. Ever since there has been a large  amount of literature \cite{fid2,x lit} about X-states. In our work, we have exploited the parametrization of X-states for different ranks \cite{fid2} to find out the comparative relations for teleportation fidelity.  
We emphasize the importance of functions of state parameters other than purity and concurrence in characterizing the nonlocal
properties of a two-qubit mixed state.
We have organized the paper as follows. In the next section, we have discussed some useful quantities related to our work. In the subsequent sections we present our results using two-qubit X-states.
% have been discussed in Section \ref{X-State}. Section \ref{con fid pur} contains the expressions of purity, concurrence and fidelity for general two-qubit X-states. In Sections \ref{2nd rank}, \ref{3rd rank} and \ref{4th rank} we have shown behaviour of the fidelity with concurrence and purity for rank two, three and four X-states respectively. 
Finally we shall conclude in the last section. 
\section{Preliminaries} \label{preli}
In this section, we introduce some relevant quantities which are required to obtain our results. Among many measures of entanglement of a two-qubit system, concurrence \cite{Wooters} is extensively used so far in many contexts. Concurrence $\mathcal{C}(\ket{\psi})$ of a pure state ${\ket{\psi}}$ is defined as $\mathcal{C}(\ket{\psi})=\langle\psi|\tilde\psi\rangle$, where $\ket{\tilde\psi}=(\sigma_y\otimes\sigma_y)\ket
{\psi^*}$. Here ($*$) is complex conjugate of $\ket{\psi}$ in computational basis and $\sigma_y$ is Pauli matrix.  
Concurrence \cite{Wooters} of a two-qubit mixed state $\rho$ is defined  as 
\begin{equation}\label{concurrence}
\mathcal{C}=\mbox{max}(0,\lambda_1-\lambda_2-\lambda_3-\lambda_4),
\end{equation}
where $\lambda_i$'s are the eigenvalues, in descending order of the matrix $\sqrt{\sqrt{\rho}\tilde{\rho}\sqrt{\rho}}$.
Here $\tilde{\rho}=(\sigma_y\otimes\sigma_y)\rho^\ast
(\sigma_y\otimes\sigma_y)$, where `$\ast$' denotes the conjugate of $\rho$. For Bell states
it yields $1$ while for separable states $C(\rho)=0$.  
%\subsubsection{Negativity}
%Negativity\cite{Vidal} of a two-qubit state $\rho$ is defined as 
%\begin{equation}\label{negativity}
%\mathcal{N}=\frac{||\rho^{T_B}||_1-1}{2},
%\end{equation}  
%where $\rho^{T_B}$ is the partial transpose of the state $\rho$ with respect to the subsystem $B$, $||\cdot||_1$ represents the trace norm and is defined as $||\Lambda||_1=\mbox{Tr}\sqrt{\Lambda^\dagger \Lambda}$.
%\subsection{Teleportation fidelity}
To define the teleportation fidelity, we shall use the prescription given by Horodecki \textit{et al.} \cite{horotele}. Maximum teleportation fidelity of a two-qubit state $\rho$ can be expressed as
\begin{equation}\label{tel fidelity}
 \mathcal{F}_{\mbox{max}}\leq \frac{1}{2}\big[1+\frac{1}{3}\mbox{Tr}\sqrt{T^\dagger T}\big],
\end{equation}
where the elements of the matrix $T$ are defined as $t_{mn}=\mbox{Tr}[\rho(\sigma_n\otimes\sigma_m)]$ and $m,n=(1,2,3)$. Furthermore, they have shown that for a state $\rho$ with $\mbox{Tr}\sqrt{T^\dagger T}>1$, the inequality in Eq. (\ref{tel fidelity}) can be replaced by a equality as \cite{horotele}
\begin{equation}\label{tel fidelity 1}
\mathcal{F}_{\mbox{max}}= \frac{1}{2}\big[1+\frac{1}{3}\mbox{Tr}\sqrt{T^\dagger T}\big].
\end{equation}
However, throughout the manuscript we will always consider the upper bound of fidelity $\mathcal{F}=\frac{1}{2}\big[1+\frac{1}{3}\mbox{Tr}\sqrt{T^\dagger T}\big]$ which is the expression of {\em optimal fidelity}.
\section{Two qubit X-state}\label{X-State}
 The parametric form of an arbitrary two-qubit X-state of a bipartite system can be represented as follow \cite{fid2}: 
%  \begin{widetext}
 \begin{equation}
   \label{eq:rho-Xm}
   \rho
	 =\left(
      \begin{matrix}
      \cos^2\theta&0&0&\sqrt xe^{i\mu}\\
      0&\sin^2\theta \cos^2\phi&\sqrt y e^{i\nu}&0\\
      0&\sqrt y e^{-i\nu}&\sin^2\theta \sin^2\phi \cos^2\psi&0\\
      \sqrt xe^{-i\mu}&0&0&\sin^2\theta \sin^2\phi \sin^2\psi
      \end{matrix}
   \right). 
\end{equation}
% \end{widetext}
with $\theta, \phi, \psi \in[0,\frac{\pi}{2}]$, $x, y \geq0$ and $\mu, \nu\in[0,2\pi]$. 
However these conditions are not enough to make Eq. (\ref{eq:rho-Xm}) a valid density matrix. 
Further constraints $x\in[0,\mathcal{H}]$ and $y\in[0,\mathcal{G}]$ are required to make it positive
semidefinite. Here we define  $\mathcal{H}=\sin^2\theta\cos^2\theta\sin^2\phi\sin^2\psi$ and 
$\mathcal{G}=\sin^4\theta\cos^2\phi\sin^2\phi\cos^2\psi$.
Mendon\c{c}a \textit{et al.} \cite{fid2} have given a parametrization of two-qubit
X-states of different ranks. A two-qubit X-state will be of rank one if $(x=\mathcal{H},y=0,\mathcal{A}=0)$ or  
$(x=0,y=\mathcal{G},\mathcal{B}=0)$, where $\mathcal{A}=\sin^2\theta(1-\sin^2\phi\sin^2\psi)$ 
and $\mathcal{B}=1-\mathcal{A}$. A rank two X-state can be parametrized as  $(x<\mathcal{H},y=0,\mathcal{A}=0)$ or 
$(x=0,y<\mathcal{G},\mathcal{B}=0)$ or $(x=\mathcal{H},y=\mathcal{G},\mathcal{AB}>0)$. Conditions
$(x<\mathcal{H},y=\mathcal{G},\mathcal{A}>0)$ or $(x=\mathcal{H},y<\mathcal{G},\mathcal{B}>0)$ 
parametrize a two-qubit X-state of rank three. A rank four X-state can be parametrized 
as $(x<\mathcal{H},y<\mathcal{G},\mathcal{AB}>0)$.
They have also given the concurrence and purity for an arbitrary X-state. We will be using those expressions along with the expression of optimal fidelity to obtain its functional relationship with  concurrence and purity. This relationship will give us the pattern how optimal fidelity changes with purity and concurrence. In addition we will find dependence on other functions of the state parameters.
We underscore the importance of these other functions to characterize the
nonlocal properties of a two-qubit mixed state.
\section{Concurrence, Optimal Fidelity and Purity for general X-states}\label{con fid pur}
In \cite{fid2}, the following expressions for purity and concurrence for an arbitrary 
X-state have been obtained,
\begin{eqnarray}
\mathcal{P}=1-2(\mathcal{A}\mathcal{B}+\mathcal{G}-y+\mathcal{H}-x)\label{purity x}\\
\mathcal{C}=2 \hspace{1mm}\text{max} \hspace{1mm}[\sqrt{x}-\sqrt{\mathcal{G}},\sqrt{y}-\sqrt{\mathcal{H}}]\label{concurrence x}
\end{eqnarray}
Using Eq. (\ref{tel fidelity}), we have evaluated the expression for optimal teleportation  fidelity for the general X-state as,
\begin{eqnarray}
\mathcal{F}&=& \frac{1}{6}\Big[3+2\sqrt{(\sqrt x+\sqrt{y})^2}+2\sqrt{(\sqrt x-\sqrt{y})^2}+\nonumber\\
&&\quad\sqrt{\Big({\cos^2\theta}-{\sin^2\theta}\big({\cos^2{\phi}}
+{\sin^2\phi}\cos 2\psi\big)\Big)^2}\Big].\label{1fidelity x}
\end{eqnarray}
Now to remove the last square root in the above expression, we write Eq. (\ref{1fidelity x}) as
\begin{eqnarray}
\mathcal{F}&=& \frac{1}{6}\Big[3+2\sqrt{(\sqrt x+\sqrt{y})^2}+2\sqrt{(\sqrt x-\sqrt{y})^2}+\nonumber\\
&&\quad\mbox{sgn}\big(k(\theta,\phi,\psi)\big)k(\theta,\phi,\psi)\Big],\label{2fidelity x}
\end{eqnarray}
where $k(\theta,\phi,\psi)=\Big({\cos^2\theta}-{\sin^2\theta}\big({\cos^2{\phi}}
+{\sin^2\phi}\cos 2\psi\big)\Big)$ and `$\mbox{sgn}$' represents the sign function.
%\begin{eqnarray}
%\mathcal{F}&=& \frac{1}{6}\Big[3+2\sqrt{(\sqrt x+\sqrt{y})^2}\nonumber\\
%&&\quad+2\sqrt{(\sqrt x-\sqrt{y})^2}\nonumber\\
%&&\quad\pm\Big({\cos^2\theta}-{\sin^2\theta}\big({\cos^2{\phi}}
%\nonumber\\&&+{\sin^2\phi}\cos 2\psi\big)\Big)\Big].\label{fidelity x}
%\end{eqnarray}
Eq. (\ref{2fidelity x}) will give two different expressions depending on the choice, $x>y$ or $y>x$. Depending on that optimal fidelity expression will involve $x$ or $y$ and will be
\begin{equation}
\mathcal{F}= \frac{1}{6}\Big[3+4\sqrt{a}+
\mbox{sgn}\big(k(\theta,\phi,\psi)\big)k(\theta,\phi,\psi)\Big],\label{general fidelity x}
\end{equation}
where $a=\mbox{max}[x,y]$.
 As stated earlier, depending upon the parameters values and ranges, we can categorize a X-state as second rank, third rank, or fourth rank state. We will deal with the case of each rank separately as the complexity of the functional relationship will grow with rank. Before starting with second rank, we note  that pure states have purity as one; for each such state, optimal teleportation fidelity and concurrence are related as,
\begin{equation}
\mathcal{F}=\frac{1}{3}(2+\mathcal{C}). 
\end{equation} 
We can see that for any pure entangled state   fidelity is larger than $\frac{2}{3}$. But it is not the case always for a mixed state, as is known, and as we will see in the subsequent section.
\section{Analysis for second rank X-states}\label{2nd rank}
\subsection{Second rank X-states of first kind}
We first consider the second rank states of first type, i.e $x<\mathcal{H},y=0,\mathcal{A}=0$. First putting $y = 0$,
in the general  expression for purity and concurrence we get,
\begin{eqnarray}
&&\mathcal{P}=1+2x-2p\sin^2\theta -2q\sin^4\theta \label{purity rank 2},\\
&&\mathcal{C}=2(\sqrt{x}-f\sin^2\theta )\label{concurrence rank 2},
\end{eqnarray}
where, 
\begin{eqnarray}
&&f=\cos\phi\sin\phi\cos\psi,
\label{0rank 2}\\
&&p=\sin^2\phi \sin^2\psi, %\hspace{2mm} 0\leq p\leq 1   
\label{1rank 2}\\
&&q=\sin^2\phi (\cos^2\phi \cos^2 \psi -\sin^2\psi).
%,\hspace{2mm} -1\leq q \leq\frac{1}{4}
\label{2rank 2}
\end{eqnarray}
And optimal fidelity is given by,
\begin{equation}
\label{fidelity rank 2}
\mathcal{F}=\frac{1}{6}\Big[3+4\sqrt{x}\pm{\cos^2\theta}\mp e{\sin^2\theta}\Big],   
\end{equation}
where,
\begin{equation}\label{3rank 2}
e={\cos^2{\phi}}
+{\sin^2\phi}\cos 2\psi.
%\hspace{2mm} -1\leq e \leq 1.
\end{equation}
and the origin of $\pm$ sign in Eq. (\ref{fidelity rank 2}) is coming due to the sign function in (\ref{2fidelity x}). 
Now, situation for the rank two state will be simpler because we also have $\mathcal{A}=\sin^2\theta(1-\sin^2\phi\sin^2\psi)=0$. That will make either $\theta=0$ or $\phi=\psi=\pi/2$. No other solutions are possible. So it will suffice to consider these two cases. But for $\theta=0$, we have $\mathcal{H}=0$. As $x<\mathcal{H}$, it should be negative (because with the other restrictions for the first kind of second rank X-states, $x$ must be less than $\mathcal{H}$, otherwise it won't be a second rank X-state). But by definition of X-states, $x$ is non-negative. So, if we allow $x$ to be at most equal to $\mathcal{H}$ (which is not allowed anyway), which is zero in this case, we just get a pure state.  So, $\theta=0$ is not a valid solution here.
Now, when $\phi=\psi=\pi/2$,
\begin{equation}\label{5rank 2}
p=1, \hspace{1mm} q=-1,\hspace{1mm}\mbox{and}\hspace{1mm} e=-1.
\end{equation}
Putting these in the expressions for purity, optimal fidelity and concurrence, we get,
\begin{eqnarray}
&&\mathcal{P}=1+2x-2\sin^2\theta+2\sin^4\theta,\label{1purity rank 2}\\
&&\mathcal{C}=2\sqrt{x}\hspace{2mm}\mbox{and}\label{1concurrence rank 2}\\
&&\mathcal{F}=\frac{1}{3}\Big[2+2\sqrt{x}\Big].\label{1fidelity rank 2}
\end{eqnarray}
Using Eq.(\ref{1concurrence rank 2}) and Eq.(\ref{1fidelity rank 2}) we obtain
\begin{equation}
\mathcal{F}=\frac{1}{3}(2+\mathcal{C}).
\end{equation}
Again, it is evident that optimal fidelity is greater than classical value whenever the state is entangled. This relation is same as that for pure states.
\subsection{Rank-2 X-states of second kind}
For this we have the parametrization, $\mathcal{B}=0,x=0,y<\mathcal{G}$. All the expressions for purity, concurrence and optimal fidelity will be same as before, just $x$ will be replaced by $y$. 
\begin{eqnarray}
&&\mathcal{P}=1+2y-2p\sin^2\theta -2qsin^4\theta, \label{2purity rank 2}\\
&&\mathcal{C}=2(\sqrt{y}-f'\sin\theta \cos\theta )\hspace{2mm}\mbox{and}\label{2concurrence rank 2}\\
&&\mathcal{F}=\frac{1}{6}\Big[3+4\sqrt{y}\pm{\cos^2\theta}\mp e{\sin^2\theta}\Big].\label{2fidelity rank 2}
\end{eqnarray}
Here, $f'=\sin\phi\sin\psi$. Now the condition $\mathcal{B}=0$, i.e $\mathcal{A}=\sin^2\theta(1-\sin^2\phi\sin^2\psi)=1$ will make either $\theta=\pi/2$ and $\phi=0$, or $\theta=\pi/2$ and $\psi=0$. No other solutions are possible.
But for the first choice, we have, $\mathcal{G}=0$. So $y$ should be negative.
So this is not a valid solution by the same arguments as above.
 Let us see what happens for the other solution i.e, $\theta=\pi/2$ and $\psi=0$. In this case, we have
\begin{equation}\label{8rank 2}
p=0, \hspace{1mm} q=\sin^2\phi\cos^2\phi,\hspace{1mm} f'=0 \hspace{1mm}\mbox{and}\hspace{1mm} e=1.
\end{equation}
We obtain,
\begin{eqnarray}
&&\mathcal{P}=1+2y-2\sin^2\phi \cos^2\phi,\label{4purity rank 2}\\
&&\mathcal{C}=2\sqrt{y}\hspace{2mm}\mbox{and}\label{4concurrence rank 2}\\
&&\mathcal{F}=\frac{1}{3}\Big[2+2\sqrt{y}\Big]\nonumber
\end{eqnarray}
So finally we have
\begin{equation}
\mathcal{F}=\frac{1}{3}(2 + \mathcal{C}).\label{4fidelity rank 2}
\end{equation}
Situation is same as before, i.e. optimal fidelity is independent of purity and we will get optimal teleportation fidelity always greater than classical value as long as the state is entangled.
\subsection{Rank-2 X-states of third kind}
This is characterized by, $x=\mathcal{H},y=\mathcal{G},0\leq A \leq 1$. The expressions change accordingly. Maximum concurrence can be $2\sqrt{x}-2\sqrt{y}$ or $2\sqrt{y}-2\sqrt{x}$. We begin with the first choice such that $x>y$. We get,
\begin{eqnarray}
&&\mathcal{P}=1+2r\sin^2\theta +2r^2\sin^4\theta,\label{5purity rank 2}\\
&&\mathcal{C}=2\sqrt{x}-2\sqrt{y}\hspace{2mm}\mbox{and}\label{5concurrence rank 2}\\
&&\mathcal{F}=\frac{1}{6}\Big[3+4\sqrt{x}\pm{\cos^2\theta}\mp e{\sin^2\theta}\Big].\label{5fidelity rank 2}
\end{eqnarray}
%or, $F=\frac{1}{6}\Big[3+4\sqrt{x}\pm{\cos^2\theta}\mp{\sin^2\theta}e\Big]$, depending upon which is greater $x$ or $y$ and,
Here,
\begin{equation}\label{9rank 2}
r=-1+\sin^2\phi\sin^2\psi. \hspace{2mm} %-1\leq r \leq 0.
\end{equation}
%We take $\sqrt{x}>\sqrt{y}$, and $F=\frac{1}{6}\Big[4+4\sqrt{x}-{\sin^2\theta}(1+e)\Big]$.
From Eq. (\ref{5purity rank 2}) we solve for $\sin^2\theta$ and get,
\begin{equation}\label{10rank 2}
\sin^2\theta=\frac{-1\pm\sqrt{2\mathcal{P}-1}}{2r}=V(\mathcal{P},\phi,\psi).
\end{equation}
 Here we choose $\mathcal{F}=\frac{1}{6}\Big[3+4\sqrt{x}+{\cos^2\theta}- e{\sin^2\theta}\Big]$. From Eq. (\ref{5concurrence rank 2}) we write, $4\sqrt{x}=2\mathcal{C}+4\sqrt{y}$ and using these,
\begin{equation}\label{6fidelity rank 2}
\mathcal{F}=\frac{1}{6}\Big[4+2\mathcal{C}+4\sqrt{y}-(1+e)V(\mathcal{P},\phi,\psi)\Big].
\end{equation}
As $e+1=-2r$, we get
\begin{equation}\label{7fidelity rank 2}
\mathcal{F}=\frac{1}{6}\Big[4+2\mathcal{C}+4\sqrt{y}+(-1\pm\sqrt{2\mathcal{P}-1})\Big].
\end{equation}
To make it optimum, we choose the plus sign and hence, $\mathcal{F}=\frac{1}{6}\Big[4+2\mathcal{C}+4\sqrt{y}+(-1+\sqrt{2\mathcal{P}-1})\Big].$ Now one can take the other sign of fidelity as well i.e., $\mathcal{F}=\frac{1}{6}\Big[3+4\sqrt{x}-{\cos^2\theta}+ e{\sin^2\theta}\Big]$.
In this case by substituting the value of $\sin^2\theta$ we get
\begin{equation}\label{8fidelity rank 2}
\mathcal{F}=\frac{1}{6}\Big[4+2\mathcal{C}+4\sqrt{y}-(-1\pm\sqrt{2\mathcal{P}-1})\Big].
\end{equation}
To make it optimum, we choose the minus sign in  the expression of $\sin^2\theta$, i.e., in Eq. (\ref{10rank 2}). Both these expressions are same but depending on the situation, we need to choose the sign of $\sin^2\theta$ properly. The plus or minus sign in optimal fidelity expression can be fixed by the sign of Eq. (\ref{10rank 2}). Therefore, without loosing generality, we can take the optimal fidelity expression as 
\begin{equation}\label{9fidelity rank 21}
\mathcal{F}=\frac{1}{6}\Big[4+2\mathcal{C}+4\sqrt{y}+(-1+\sqrt{2\mathcal{P}-1})\Big].
\end{equation}
We will encounter similar kind of situation for other ranks of X-states as well. By giving similar argument and without loosing generality we can take optimal fidelity as 
\begin{equation}\label{9fidelity rank 2}
\mathcal{F}
%=\frac{1}{6}\Big[3+4\sqrt{a}+\cos^2\theta-e\sin^2\theta\Big]
=\frac{1}{6}\Big[4+4\sqrt{a}-(1+e)\sin^2\theta\Big],
\end{equation}
where, $a=\mbox{max}[x,y]$ and we will consider accordingly the sign in the expression of $\sin^2\theta$ (as in Eq. (\ref{10rank 2})) to make fidelity optimal. We will use this expression throughout the manuscript. 
Now, as the minimum value of $(1+e)$ can be zero, in the expression for $\sin^2\theta$ we have to choose the minus sign for the optimum fidelity. For this choice, it is evident from Eq.(\ref{10rank 2}) that $\sin^2\theta=V(P,\phi,\psi)$, decreases as $P$ increases for any $\phi,\psi$. Hence from Eq.(\ref{9fidelity rank 21}) optimal fidelity will also increase as the purity increases keeping concurrence and other parameters fixed at any values. Also the optimal fidelity changes monotonically with respect to parameters other than purity, or concurrence, here $y$. This is one of the main message of this paper and as we will show in the following that the same conclusion holds for 3rd and 4th rank X-states also.
%We can now set the values for $\sqrt{y}$ and see the variations of fidelity with concurrence for fixed purity, or fidelity with purity for fixed concurrence. 
The expression of $V(P,\phi,\psi)$ shows a very interesting feature. As fidelity is always a real quantity, we must have $\mathcal{P}\geq 1/2$. So, this physical constraint also restricts the minimum purity a second rank X-state can have. This fact is also evident from the expression of purity, i.e., Eqn. (\ref{5purity rank 2}). It can be shown that minimum value that $\mathcal{P}$ can take is $1/2$.
As stated earlier, from Eqn. (\ref{6fidelity rank 2}) and   (\ref{7fidelity rank 2}), it is evident that for a fixed value of $y$ and $\mathcal{C}$, optimal fidelity increases with the increment of purity. We emphasize this fact by plotting optimal fidelity with respect to purity for $y=0.01$ and $\mathcal{C}=0.2$. 
%This example shows our conclusion in a more lucid way.

\begin{figure}[h]
\centering
\includegraphics[scale=0.65]{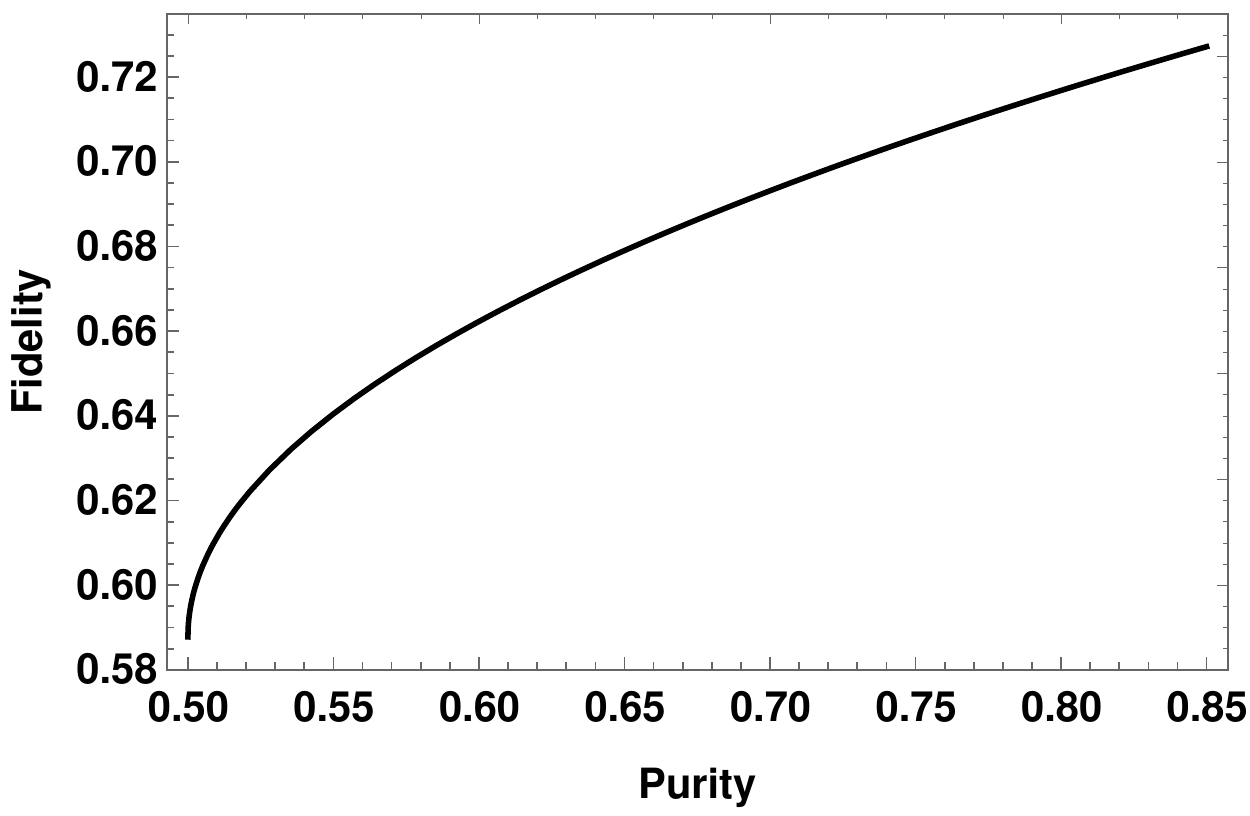}
\caption{Variation of optimal fidelity with purity for $y=0.01$ and $\mathcal{C}=0.2$ of second rank X-state of third kind.}
\label{1_fig_rank_2_x}
\end{figure}
\begin{figure}[h]
\centering
\includegraphics[scale=0.65]{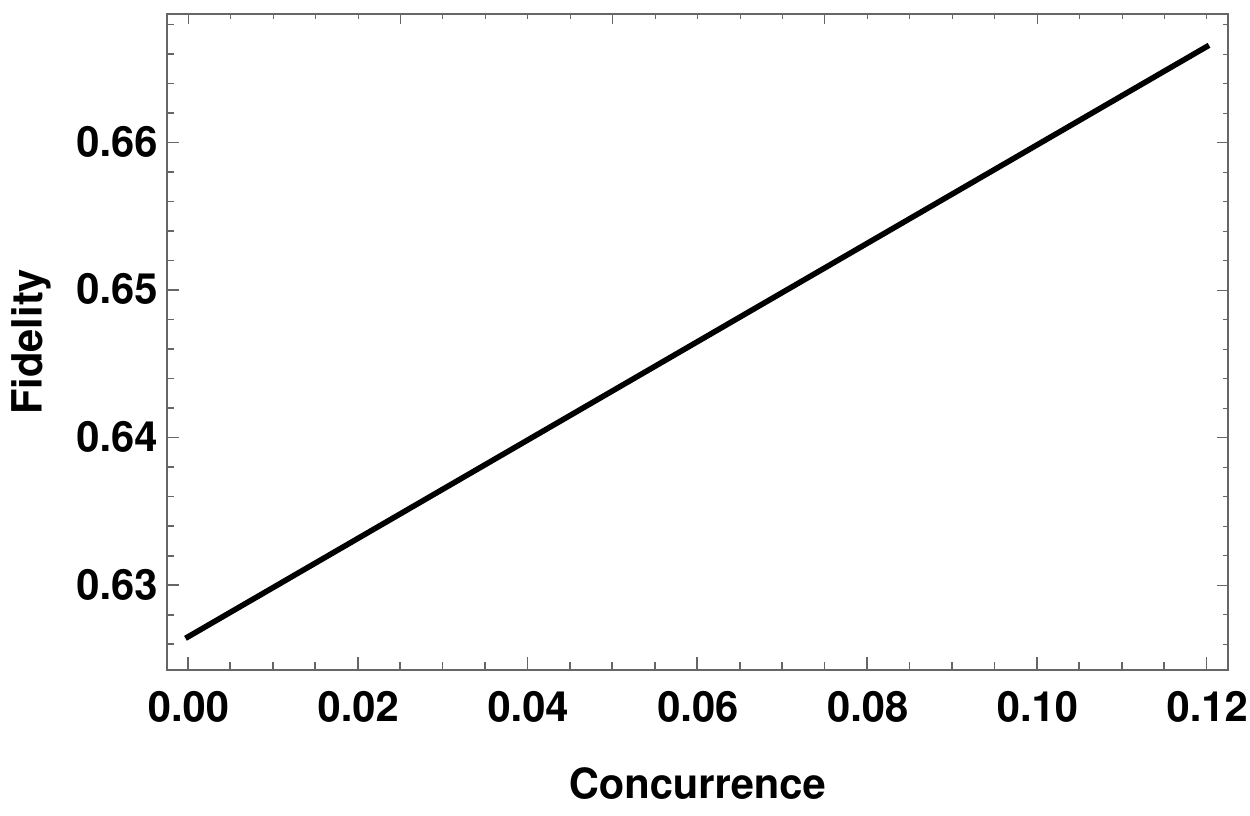}
\caption{Variation of optimal fidelity with concurrence for $y=0.01$ and $\mathcal{P}=0.7$ of second rank X-state of third kind.}
\label{2_fig_rank_2_x}
\end{figure}
It is clear from FIG. \ref{1_fig_rank_2_x}, that if a state  is not suitable for teleportation, by increasing its purity, one can make it effective for teleportation. Therefore, optimal fidelity not only depends on entanglement but also on the purity of the state. In FIG. \ref{2_fig_rank_2_x} we have plotted optimal fidelity for a fixed value of purity to show its variation with entanglement of the state and indeed optimal fidelity is increasing with concurrence. 
Here we emphasize that it may happen a state with less entanglement but higher value of purity can achieve higher optimal fidelity. We see it explicitly in the  following example. Let us consider a state with $\mathcal{P}=0.6$, $\mathcal{C}=0.2$ and $y=0.001$. For this state $\theta\approx 0.5809$, $\phi\approx 0.3124$ and $\psi\approx 1.2036$. Optimal fidelity for this state is $\mathcal{F}\approx 0.6623$. Let us take another state with $\mathcal{P}=0.7$, $\mathcal{C}=0.15$ and $y=0.001$. For this state, we have  $\theta\approx 1.4515$, $\phi\approx 1.4953$ and $\psi\approx 1.1304$. In this case optimal fidelity is $\mathcal{F}\approx 0.6765$. So a state that has less entanglement but more purity can provide higher optimal fidelity.  
Another parameter on which optimal fidelity depends on is $y$. From the optimal fidelity expression, it is clear
that it increases monotonically with $y$. This is the first example where we see the dependence of optimal fidelity on
properties other than purity and concurrence. This parameter also seems to characterize the nonlocal
properties of the state.  \\~\\
Now if we consider the second choice of concurrence i.e., $\mathcal{C}=2\sqrt{y}-2\sqrt{x}$ when $y>x$. Everything will remain same except $x$ in Eq. (\ref{5fidelity rank 2}) will change to $y$ and $y$ in Eq.(\ref{6fidelity rank 2}) and (\ref{7fidelity rank 2}) 
will change to $x$. All the arguments and results remain the same.
\section{Analysis for Third rank X-states}\label{3rd rank}
\subsection{Rank-3 X-states of first kind}
\begin{figure}[h]
\centering
\includegraphics[scale=0.65]{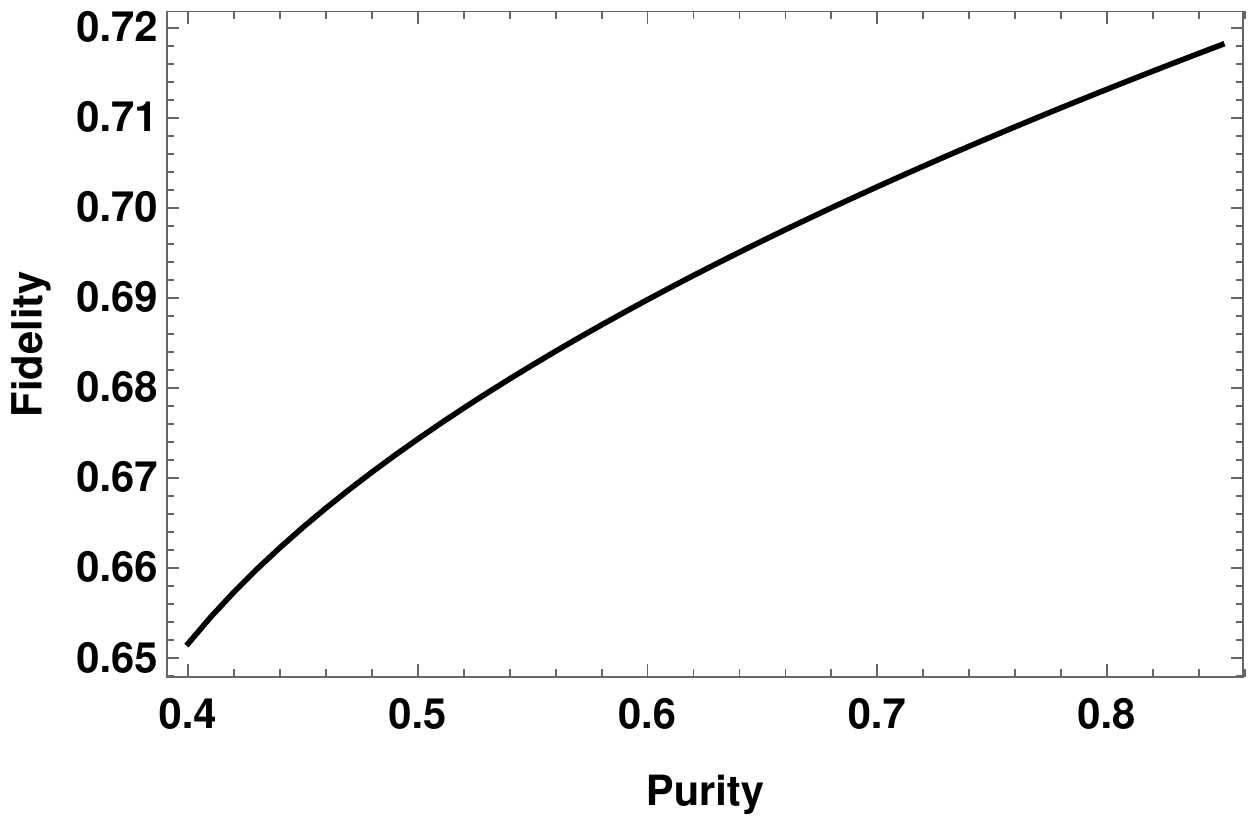}
\caption{Variation of optimal fidelity with purity for $d=\frac{3}{4}$, $e=0$ and $f=0$ with $\mathcal{C}=0.2$ of third rank X-state of first kind.}
\label{1fig_rank_3_x}
\end{figure}

\begin{figure}[h]
\centering
\includegraphics[scale=0.65]{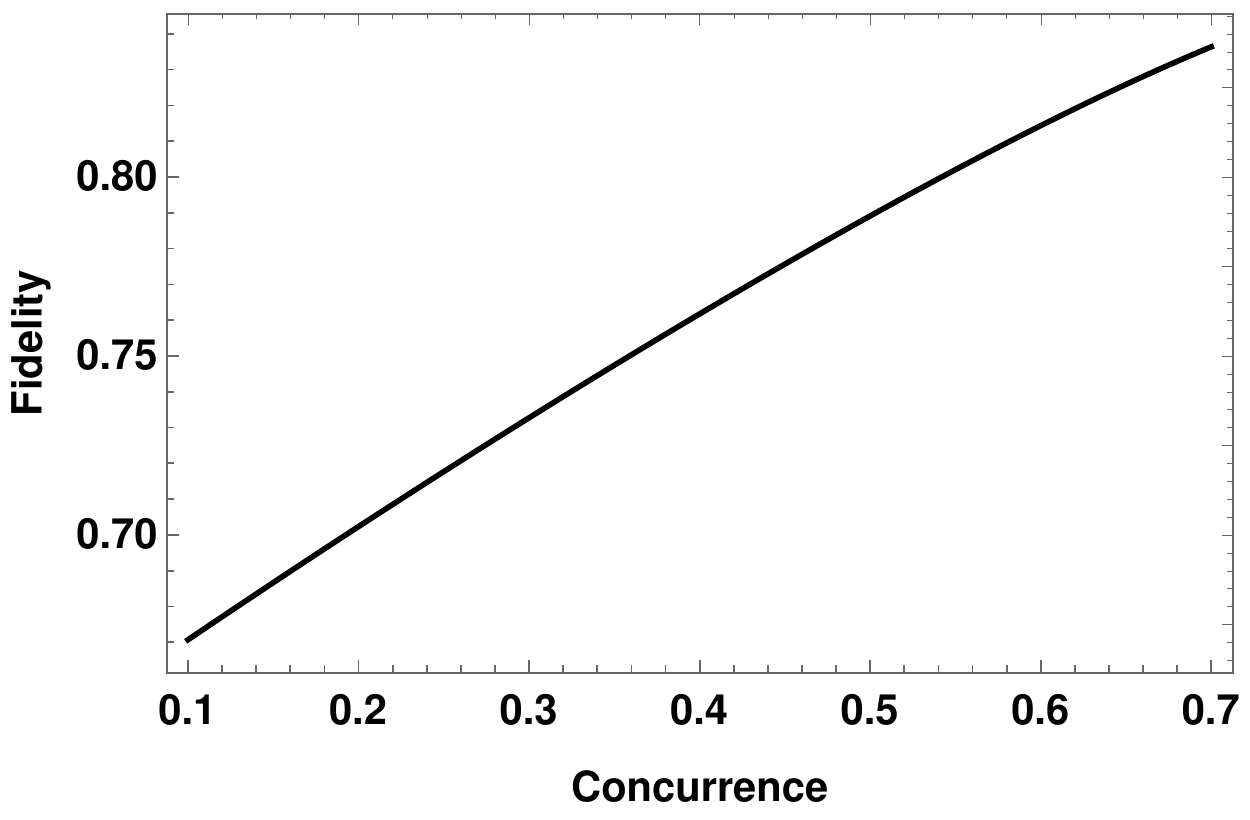}
\caption{Variation of optimal fidelity with concurrence for $d=\frac{3}{4}$, $e=0$ and $f=0$ with $\mathcal{P}=0.7$ of third rank X-state of first kind.}
\label{2fig_rank_3_x}
\end{figure}

\begin{figure}[h]
\centering
\includegraphics[scale=0.65]{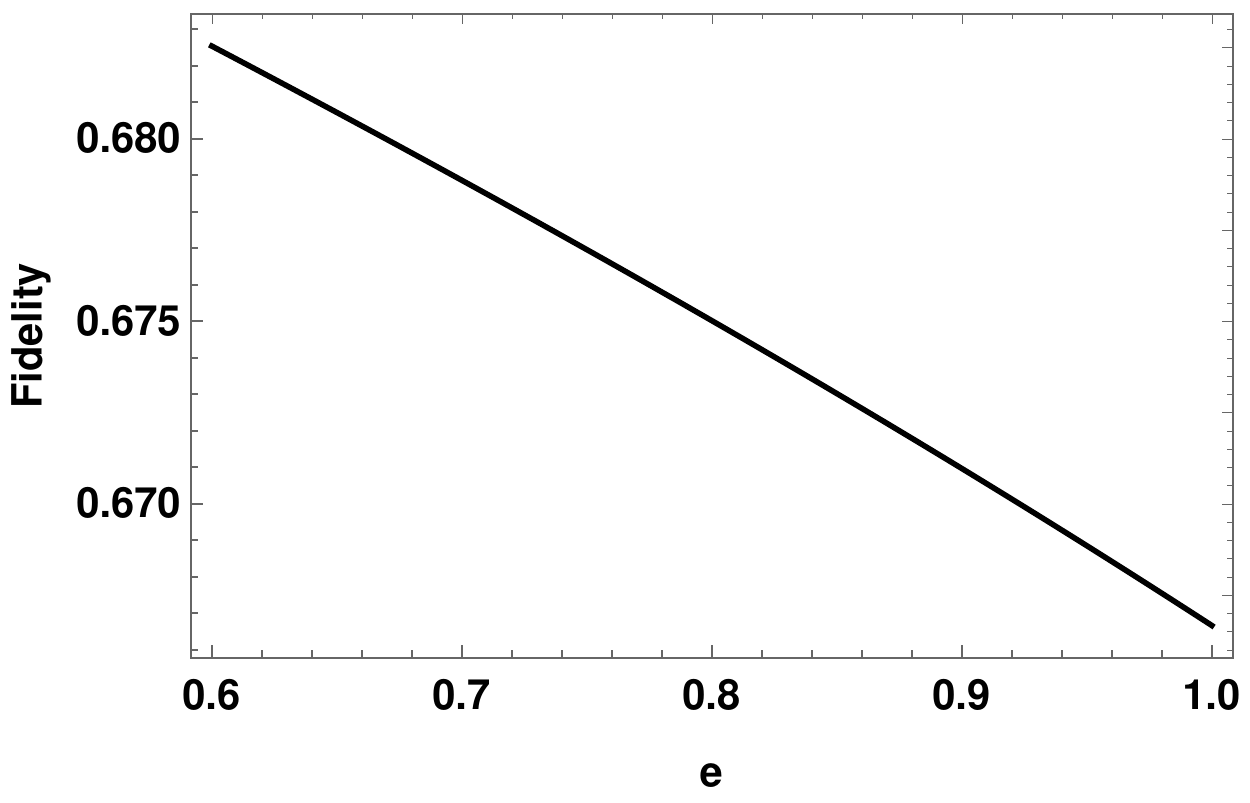}
\caption{Variation of optimal fidelity with '\textbf{e}' for $f=0$, $\mathcal{P}=.7$ and $\mathcal{C}=0.2$ of third rank X-state of first kind.}
\label{fig_e}
\end{figure}

\begin{figure}[h]
\centering
\includegraphics[scale=0.65]{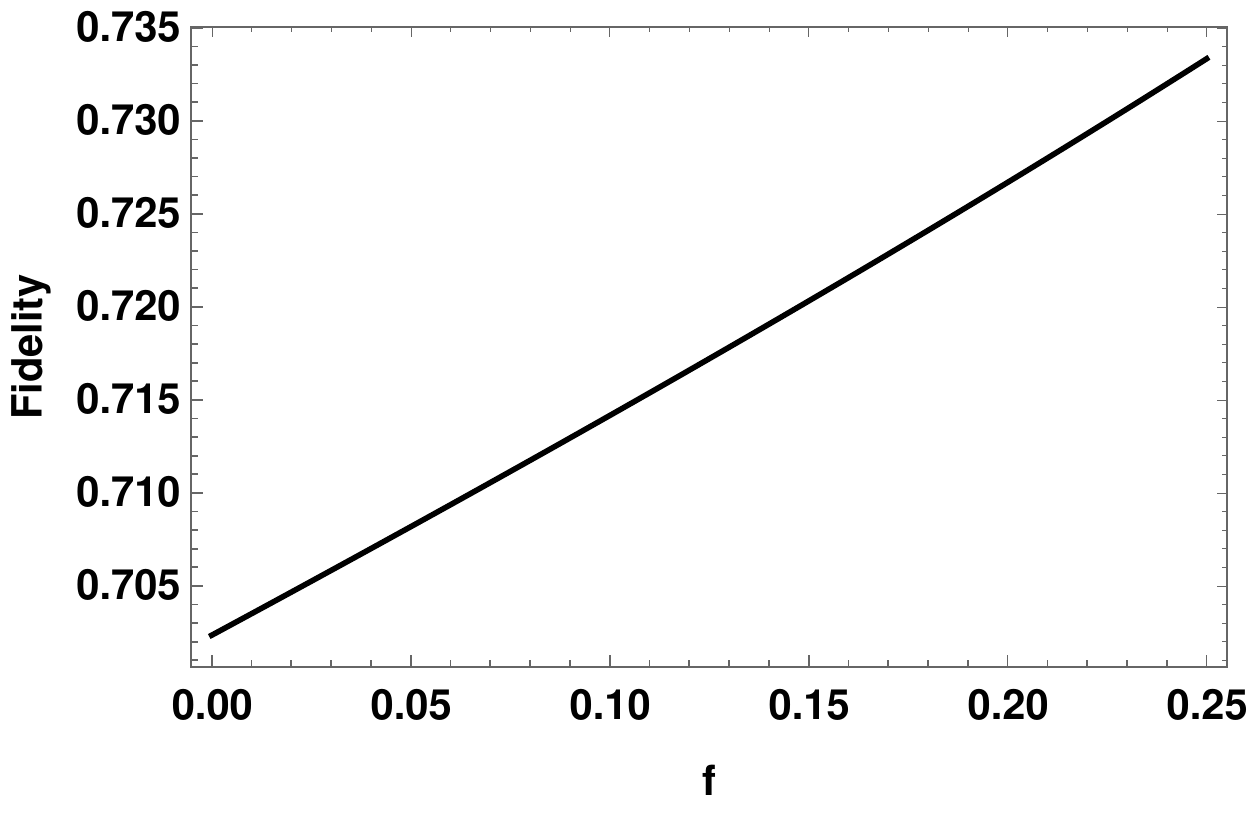}
\caption{Variation of optimal fidelity with '\textbf{f}' for $e=0$, $\mathcal{P}=.7$ and $\mathcal{C}=0.2$ of third rank X-state of first kind.}
\label{fig_f}
\end{figure}

Third rank X-state of first kind is characterized by, $x<\mathcal{H}, y=\mathcal{G},A>0$. Given that we get,
\begin{eqnarray}
&&\mathcal{P}=1+2x-2 \sin^2\theta+2d\sin^4\theta,\hspace{2mm} \label{purity rank 3}\\
&&\mathcal{C}=2\hspace{1mm}\text{max}\hspace{1mm}[\sqrt{x}-f \sin^2\theta , \sqrt{y}- f'\sin\theta \cos\theta ]\hspace{1mm}\nonumber\\
&&\label{concurrence rank 3}\\
&&\mathcal{F}=\frac{1}{6}\Big[3+4\sqrt{a}+{\cos^2\theta}-{e\sin^2\theta}\Big],
\label{fidelity rank 3}\end{eqnarray}
%where we are taking that $x>y$ and  
where, $e,f,f',a$ are same as before and,
\begin{equation}\label{rank 3}
d=1-\sin^2\phi\sin^2\psi+\sin^4\phi\sin^4\psi.
%,\hspace{2mm} 3/4\leq d\leq 1.
\end{equation}
%To be restricted in the class of first kind, we have to be careful about the lower and upper bound of the parameters $d,e,f$ and $f'$. 
We first take $x>y$. As $x>y$ implies $x>\mathcal{G}$ and $x<\mathcal{H}$, we have $\mathcal{H}>x>\mathcal{G}$ and 
for this choice, $\mathcal{C}$ is $2(\sqrt{x}-f\sin^2\theta)$ and $\mathcal{F}$ is  
$\frac{1}{6}\Big[3+4\sqrt{x}+{\cos^2\theta}-{e\sin^2\theta}\Big].$
Now, solving for $\sin^2\theta$ from Eq. (\ref{purity rank 3}) and (\ref{concurrence rank 3}) we get, 
%\begin{equation}
%
%\sin^2\theta=
%
%\frac{(1-f\mathcal{C})\pm\sqrt{(1-f\mathcal{C})^2-
%(1-\mathcal{P}+\mathcal{C}^2/2)(2d+2f^2)}}{(2d+2f^2)}.
%
%\end{equation}
\begin{equation}
\begin{split}
\sin^2\theta &=V_1(\mathcal{P},\mathcal{C},\phi,\psi)\\
&=\frac{(1-f\mathcal{C})\pm\sqrt{(1-f\mathcal{C})^2-(1-\mathcal{P}+\mathcal{C}^2/2)(2d+2f^2)}}{(2d+2f^2)}
\end{split}
\end{equation}
Using the expression for $\mathcal{C}$ and the evaluated $\sin^2\theta$, we now write optimal fidelity $\mathcal{F}$ in
terms of $\mathcal{C}, \mathcal{P}, d, e$ and $f$ as,
\begin{equation}\label{1fidelity rank 3}
\mathcal{F}=\frac{1}{6}\big[4+2\mathcal{C}-(1+e-4f){V_1}(\mathcal{P},\mathcal{C},\phi,\psi)\big].
\end{equation}
As stated for the 2nd rank case, to get optimum fidelity, we have to choose the minus sign of 
$V_1(\mathcal{P},\mathcal{C},\phi,\psi)$ as the minimum value of $(1+e-4f)$ can be zero. Then it is evident from the expression that for any values  of $\phi,\psi$ and $\mathcal{C}$, $\mathcal{F}$ increases with $\mathcal{P}$, as $V_1(\mathcal{P},\mathcal{C},\phi,\psi)$ decreases with the increase of $P$. So as before the same result holds for 3rd rank X-states of first kind. 
To illustrate this behavior graphically, we set $\phi=\frac{\pi}{4}$, $\psi=\frac{\pi}{2}$ and then the 
Eq. (\ref{1fidelity rank 3}) reduces to the following form,
%$d=\frac{3}{4}$, $e=0$ and $f=0$. For these values of $d,e$ and $f$ two sets of $\phi$ and $\psi$ are possible and these are $\phi=\frac{\pi}{4}$, $\psi=\frac{\pi}{2}$ and $\phi=\frac{\pi}{2}$, $\psi=\frac{\pi}{4}$. We choose the first solution and then the Eq. (\ref{1fidelity rank 3}) reduces to the following form
\begin{equation}\label{2fidelity rank 3}
\mathcal{F}=\frac{1}{18}\bigg(10+6\mathcal{C}+\sqrt{6\mathcal{P}-
3\mathcal{C}^2-2}\bigg).
\end{equation}
We plot this expression for optimal fidelity $\mathcal{F}$ with purity $\mathcal{P}$ for a fixed entanglement, i.e concurrence $\mathcal{C}$.
FIG. \ref{1fig_rank_3_x}, shows that for a fixed entanglement, the optimal fidelity
increases with purity. Like second rank X-state, optimal fidelity of these states also increases with concurrence for fixed purity as shown in FIG. \ref{2fig_rank_3_x}.
% As purity has non trivial effect on fidelity, we may expect that there exist a minimum of purity below which a mixed 
%entangled state is not found to produce fidelity  better than the classical value. We have found numerically the minimum purity
% $\approx 0.38$ for this particular class of states below which it ceases to behave as a quantum teleportation resource. In this situation increase of concurrence would not help to increase its fidelity.
 Moreover the right hand side of Eq.(\ref{1fidelity rank 3}) involves few more parameters. Here we are giving a very interesting example. Consider a state with $\mathcal{P}=0.6$, $\mathcal{C}=0.2$, $\phi=\frac{\pi}{4}$ and $\psi=\frac{\pi}{2}$. For this state, we have $\mathcal{F}\approx 0.6898$. These values of $\psi$ and $\phi$ give $d=\frac{3}{4}$, $e=0$ and $f=0$. For another state with $\mathcal{P}=0.64$, $\mathcal{C}=0.22$, $\phi=\frac{\pi}{2}$ and $\psi=\frac{2\pi}{25}$, we get $\mathcal{F}\approx 0.6612$. For this state $d \approx 0.9419$, $e=0.87563$ and $f=0$. So with less entanglement and purity one can have more optimal fidelity for different values of $\phi$ and $\psi$ or $d$, $e$ and $f$. 
Now, let us see the ranges of $d$, $e$ and $f$. We obtain
\begin{eqnarray}
\frac{3}{4}\leq d <1\\
-1<e<1\\
0\leq f <\frac{1}{2}
\end{eqnarray}
The ranges of $d$ , $e$ and $f$ have been obtained by maximizing and minimizing the functions independently. However, as they all are functions of $\phi$ and $\psi$, they can not be varied independently.
As  optimal fidelity depends on those parameters also rather than only depending on purity and concurrence, we have plotted variations of optimal fidelity with those parameters in Fig.(\ref{fig_e}) and Fig.(\ref{fig_f}) showing optimal fidelity decreases monotonically with $e$, whereas it increases monotonically with $f$. In the figures the ranges of the parameters $e$ and $f$ have been appropriately modified.
Now, we are left with the situation when $y>x$. In this case the concurrence will be $2(\sqrt{y}-f'\sin\theta\cos\theta)$ and $\mathcal{F}=\frac{1}{6}\Big[3+4\sqrt{y}+{\cos^2\theta}-{e\sin^2\theta}\Big]$. Doing similar kind of calculation one can show that 
\begin{eqnarray}\label{3fidelity rank 3}
\mathcal{F}=&&\frac{1}{6}\Big[4+2\mathcal{C}+4f'\sqrt{V_2(\mathcal{P},x,\phi,\psi)(1-V_2(\mathcal{P},x,\phi,\psi))}\nonumber\\
&&-(1+e)V_2(\mathcal{P},x,\phi,\psi)\Big],
\end{eqnarray} 
where $V_2(\mathcal{P},x,\phi,\psi)=\frac{1\pm\sqrt{1-2d(1+2x-\mathcal{P})}}{2d}$. Here also one can easily verify a similar kind of trend as before. 
\subsection{Rank-3 X-states of second kind}
This class of states are characterized by $x=\mathcal{H}$, $y<\mathcal{G}$, $\mathcal{A}<1$. We get,
\begin{eqnarray}
&&\mathcal{P}=1+2y-2t \sin^2\theta +2 u \sin^4\theta, \label{1purity rank 3}\\
&&\mathcal{C}=2\hspace{1mm}\text{max}\hspace{1mm}[\sqrt{x}- f\sin^2\theta, \sqrt{y}- f'\sin\theta \cos\theta ]\hspace{1mm}\nonumber\\
&&\label{1concurrence rank 3}\\
&&\mathcal{F}=\frac{1}{6}\Big[3+4\sqrt{a}+{\cos^2\theta}- e {\sin^2\theta}\Big] \label{4fidelity rank 3},
\end{eqnarray}
where, $t=1-p$. $p,a$ are same as before and,
\begin{equation}
u=1+\sin^4\phi\sin^4\psi-\cos^2\phi\cos^2\psi\sin^2\phi-2\sin^2\phi\sin^2\psi.
\end{equation}
First we take $x>y$. As $x>y$ implies $\mathcal{H}>y$ and $y<G$, concurrence will be $2(\sqrt{x}-f\sin^2\theta)$ and $\mathcal{F}$ is  $\frac{1}{6}\Big[3+4\sqrt{x}+{\cos^2\theta}-{e\sin^2\theta}\Big]$. After doing a calculation as above, we get
\begin{equation}\label{5fidelity rank 3}
\mathcal{F}=\frac{1}{6}\Big[4+2\mathcal{C}-(1+e-4f)V_3(\mathcal{P},\phi,\psi,y)\Big],
\end{equation}   
where $V_3(\mathcal{P},\phi,\psi,y)=\frac{t\pm\sqrt{t^2-2u(1+2y-\mathcal{P})}}{2u}$. So, the situation is similar as first kind and we would be getting similar results.
Now, we will consider the situation $y>x$ i.e., $\mathcal{G}>y>\mathcal{H}$. Here concurrence will be $2(\sqrt{y}+\sin\theta\cos\theta f')$ and $\mathcal{F}$ to be $\frac{1}{6}\Big[3+4\sqrt{y}+{\cos^2\theta}-e{\sin^2\theta}\Big]$. In this situation calculation will be slightly different. The reason is that now the expression for purity $\mathcal{P}$ involves $y$, not $x$. So, in this case we will be getting a 4th order equation of $\sin^2\theta$ from the expression of $\mathcal{P}$ and $\mathcal{C}$. We will not do this in this section as in the next section for 4th rank
 X-states we will discuss a similar situation.
\section{Analysis for Rank-4 X-states}\label{4th rank}
\begin{figure}[h]
\centering
\includegraphics[scale=0.65]{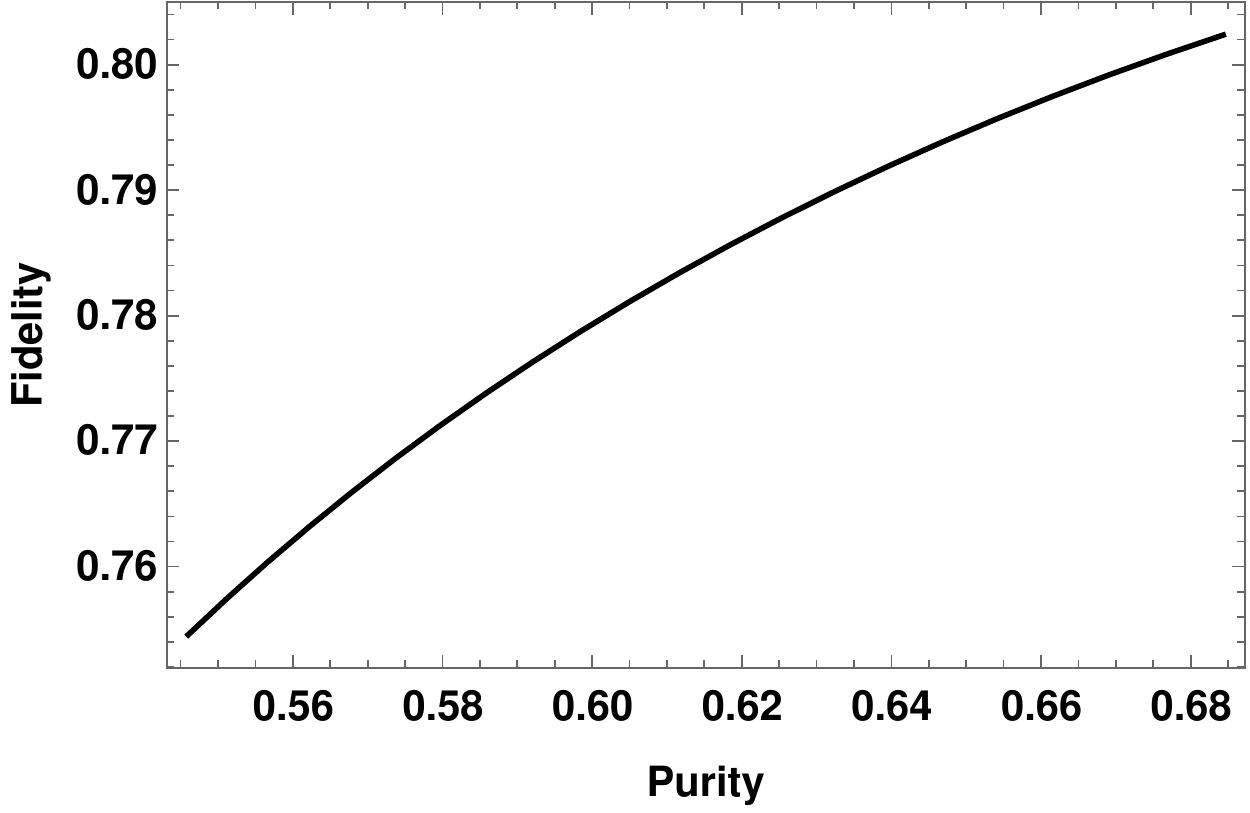}
\caption{Variation of optimal fidelity with purity for $g=\frac{11}{16}$, $e=\frac{1}{2}$, $f=\frac{1}{2\sqrt{2}}$ and $y=0$ with $\mathcal{C}=0.2$ of rank-4 X-state.}
\label{fig_rank_4_x}
\end{figure}

\begin{figure}[h]
\centering
\includegraphics[scale=0.65]{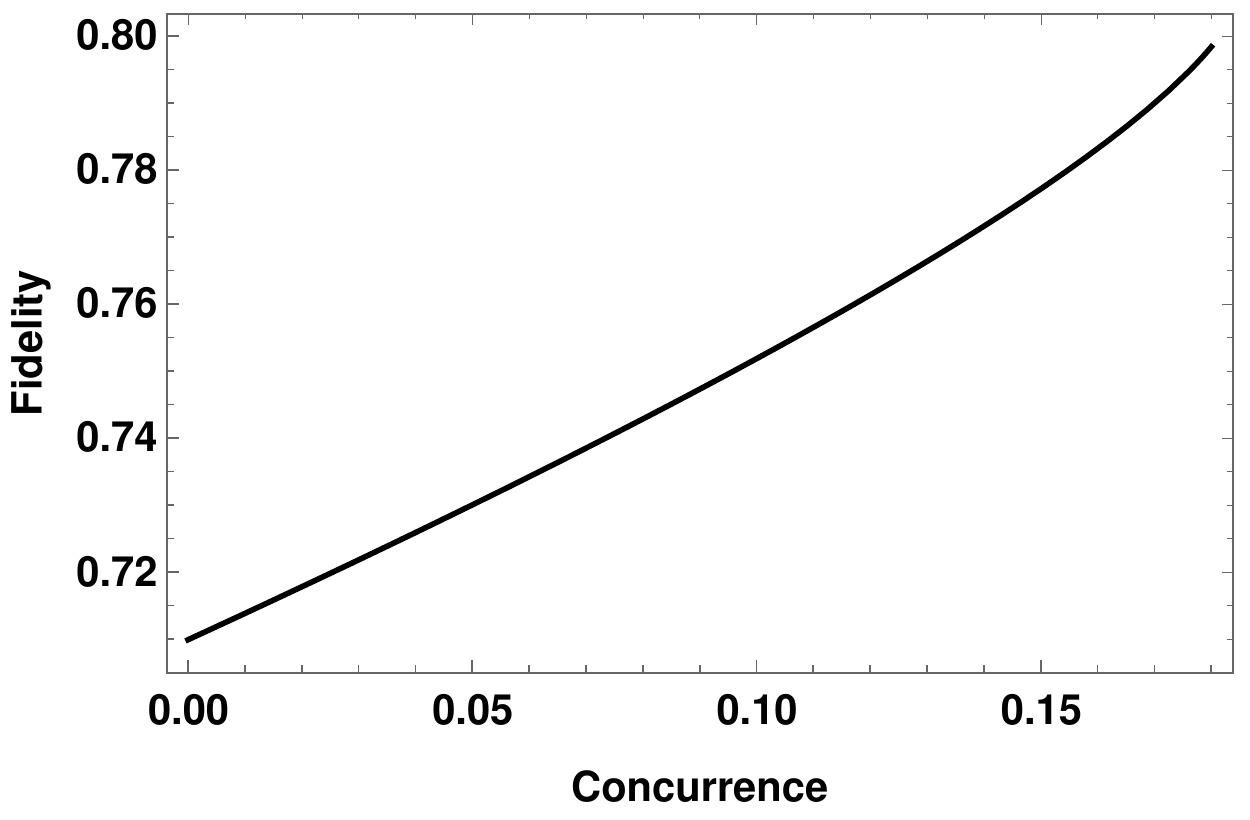}
\caption{Variation of optimal fidelity with purity for $g=\frac{11}{16}$, $e=0$, $f=\frac{1}{2\sqrt{2}}$ and $y=0$ with $\mathcal{P}=0.7$ of rank-4 X-state.}
\label{2fig_rank_4_2_x}
\end{figure}

\begin{figure}[h]
\centering
\includegraphics[scale=0.65]{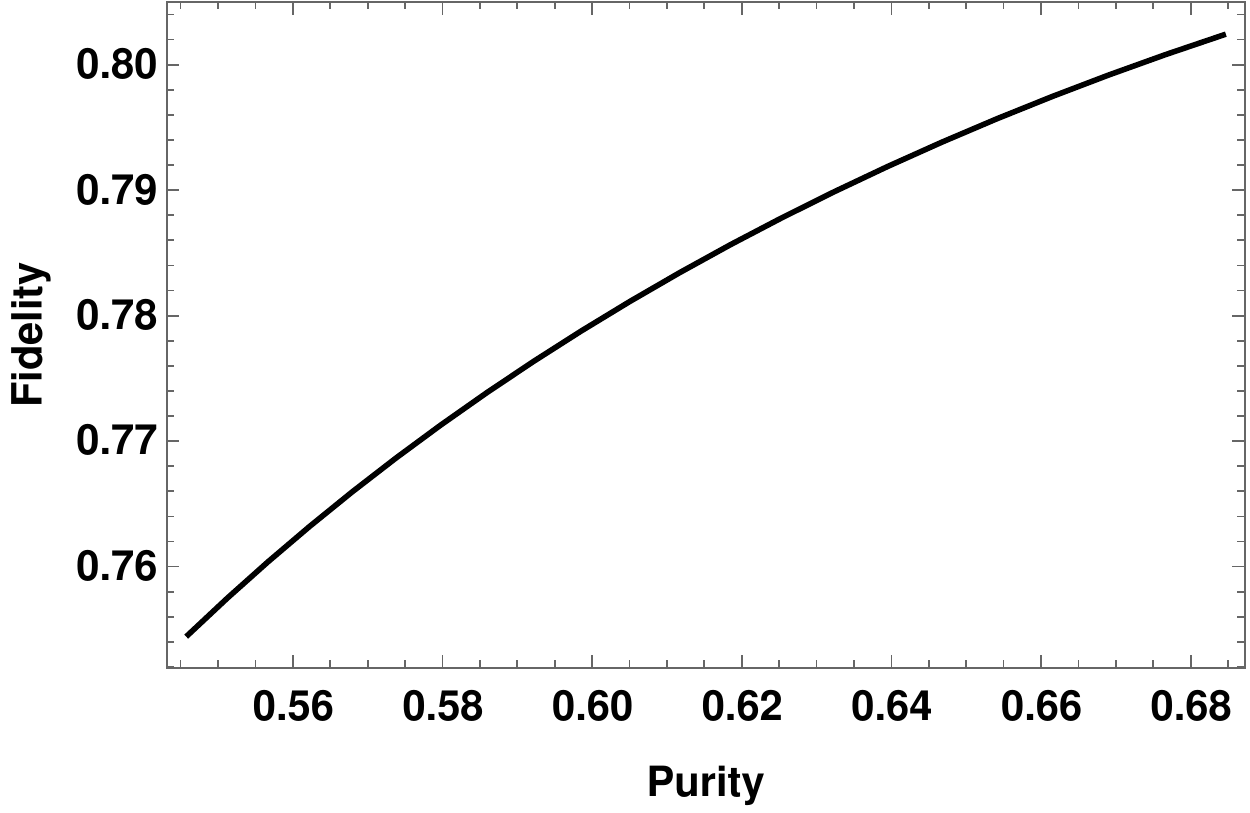}
\caption{Variation of optimal fidelity with purity for $g=\frac{11}{16}$, $f'=\frac{1}{2}$ and $x=0$ with $\mathcal{C}=0.2$ of fourth rank X-state.}
\label{1fig_rank_4_2_x}
\end{figure}

General fourth rank X-states will be characterized by $x<\mathcal{H}$, $y<\mathcal{G}$, $\mathcal{A}\mathcal{B}>0$.
Putting the values of $\mathcal{A},\mathcal{B},\mathcal{G}, \mathcal{H}$ we get the values of $P$ and $C$ as,
\begin{eqnarray}
&&\mathcal{P}=1+2x+2y-2 \sin^2\theta+2\sin^4\theta g, \label{purity rank 4}\\
&&\mathcal{C}=2\hspace{1mm}\text{max}[\sqrt{x}-  f\sin^2\theta, \sqrt{y}- f'\sin\theta \cos\theta ]\hspace{1mm}\nonumber\\
&&\label{concurrence rank 4}\\
&&\mathcal{F}=\frac{1}{6}\Big[3+4\sqrt{a}+{\cos^2\theta}-e{\sin^2\theta}\Big],
 \label{fidelity rank 4}
 \end{eqnarray}
where,
\begin{eqnarray}
&&g=\frac{1}{64}[53+4\cos 2\phi +7\cos 4\phi +8\cos 4\psi \sin^4\phi],\nonumber\\
&&\label{rank 4}\\
&&f=\sqrt{\sin^2\phi \cos^2 \phi \cos^2\psi}\hspace{2mm}\mbox{and}\label{1rank 4}\\
&&f'=\sqrt{\sin^2\phi \sin^2\psi}.\label{2rank 4}
\end{eqnarray}
%The ranges of these functions are given as,
%\begin{eqnarray}
%0< f < \frac{1}{2}\\
%0< f' < \frac{1}{2}\\
%\frac{2}{3}\leq g < 1
%\end{eqnarray}
%So, again among these three only two are independent and we can write the remaining parameter as a function of the independent ones. 
First, we choose $x>y$ and also $\sqrt{x}-  f\sin^2\theta> \sqrt{y}- f'\sin\theta \cos\theta$. So, we take $\mathcal{C}$ to be $2(\sqrt{x}-f\sin^2\theta)$ and $\mathcal{F}$ to be $\frac{1}{6}\Big[3+4\sqrt{x}+{\cos^2\theta}-e{\sin^2\theta}\Big]$.
Now, from Eq. (\ref{purity rank 4}) and (\ref{concurrence rank 4}) we get, 
\begin{eqnarray}
&&\sin^2\theta=V_4(\mathcal{P},\mathcal{C},\phi,\psi,y)=\nonumber\\
&&\frac{(1-f\mathcal{C})\pm\sqrt{(1-f\mathcal{C})^2-(1- \mathcal{P}+2y+\mathcal{C}^2/2)(2g+2f^2)}}{(2g+2f^2)}.\nonumber\\
\end{eqnarray}
Using the expression for $\mathcal{C}$ and the evaluated $\sin^2\theta$, we now write optimal fidelity $\mathcal{F}$ in terms of $\mathcal{C}, \mathcal{P}, \phi,\psi$ and $y$ as,
\begin{equation}\label{1fidelity rank 4}
\mathcal{F}=\frac{1}{6}\big[4+2\mathcal{C}-(1+e-4f)V_4(\mathcal{P},\mathcal{C},\phi,\psi,y)\big].
\end{equation}
In similar fashion here also we can argue that as purity increases keeping others constant, optimal fidelity also increases and also it is evident from the FIG. \ref{fig_rank_4_x}. For this plot we choose $\psi=\phi=\frac{\pi}{4}$ or $\psi=2\tan^{-1}(\sqrt{5-2\sqrt{6}})$, $\phi=\frac{\pi}{3}$ and $y=0$. For these values of $\phi$ and $\psi$, $g=\frac{11}{16}$, $f=\frac{1}{2\sqrt{2}}$ and $e=\frac{1}{2}$. FIG. \ref{2fig_rank_4_2_x} shows the variation of optimal fidelity with concurrence for a fixed value of purity.
Now for $x>y$, we could have $\sqrt{x}-  f\sin^2\theta< \sqrt{y}- f'\sin\theta \cos\theta$. So $\mathcal{C}=2(\sqrt{y}- f'\sin\theta \cos\theta)$. In this case from the expression of $\mathcal{P}$ and $\mathcal{C}$ we will get a fourth order equation for $\sin^2\theta$, which will not involve $y$ In principle we will get four solutions of $\sin^2\theta$ from this equation as a function of $\phi,\psi$ and $x$. Putting these solutions of $\sin^2\theta$ in the expression for $\mathcal{F}$, we will have $\mathcal{F}$ as a function of $C,P,\phi,\psi$ and $x$. The fourth order equation will be similar like we will derive in the following for the case of $y>x$ and $\sqrt{x}-  f\sin^2\theta< \sqrt{y}- f'\sin\theta \cos\theta$.
%We will show similar kind of situation in the case for $y>x$. 
So, let us consider the case when $y>x$ and $\sqrt{x}-  f\sin^2\theta< \sqrt{y}- f'\sin\theta \cos\theta$, we have the value of concurrence to be $2(\sqrt{y}-f'\sin\theta\cos\theta )$. In this case we need to replace $x$ by $y$ in the optimal fidelity expression given in Eq. (\ref{fidelity rank 4}). Here also we will get a fourth order equation of $\sin^2\theta$. Using the expression for purity $\mathcal{P}$ and concurrence $\mathcal{C}$, we get the following equation,
\begin{eqnarray}
&&\alpha^2\sin^8\theta+2\alpha\beta \sin^6\theta +[\beta^2+2\alpha(1+2x+\mathcal{C}^2/2-\mathcal{P})\nonumber\\
&&+4\mathcal{C}^2f'^2]\sin^4\theta
+[2\beta(1+2x+\mathcal{C}^2/2-\mathcal{P})-4C^2f'2]
\nonumber\\
&&\sin^2\theta
+(1+2x+\mathcal{C}^2/2-\mathcal{P})^2=0,\label{3rank 4}
\end{eqnarray}
where,
\begin{equation}\label{4rank 4}
\alpha=2g-2f'^2 \hspace{2mm}\mbox{and}\hspace{2mm} \beta=2f'^2-2.
\end{equation}
From this equation, in principle one can get four solutions for $\sin^2\theta$ and using that one can get the expression for optimal fidelity $\mathcal{F}$ in terms of $\mathcal{C}$, $\mathcal{P}$, $x$, $\phi$ and $\psi$. 
As solving this equation will be very involved, we avoid that and get some plots for some particular values of the parameters showing the pattern. We choose $x=0$, $\phi=\psi=\frac{\pi}{4}$ or $\phi=\frac{\pi}{3}$, $\psi=-2\tan^{-1}(\sqrt{2}-\sqrt{3})$ and $\mathcal{C}=0.2$.  In FIG. \ref{1fig_rank_4_2_x}, we see a trend as before, i.e, optimal fidelity increases with purity for a fixed concurrence.
Now finally we are left with $y>x$ and $\sqrt{x}-  f\sin^2\theta> \sqrt{y}- f'\sin\theta \cos\theta$. In this case $\mathcal{C}=2(\sqrt{x}-  f\sin^2\theta)$ and $\mathcal{F}=\frac{1}{6}\Big[3+4\sqrt{y}+{\cos^2\theta}-e{\sin^2\theta}\Big]$. After few steps of calculation, we find
\begin{equation}\label{2fidelity rank 4}
\mathcal{F}=\frac{1}{6}\Big[4+4\sqrt{y}-(1+e)V_5(\mathcal{P},\mathcal{C},y,\phi,\psi)\Big],
\end{equation}     
where, 
\begin{eqnarray}\label{5rank 4}
&&V_5(\mathcal{P},\mathcal{C},y,\phi,\psi)=\sin^2\theta=\\
&&\frac{(1-f\mathcal{C})\pm\sqrt{(1-f\mathcal{C})^2-(2g+2f^2)(1+\mathcal{C}^2/2+2y-\mathcal{P})}}{(2g+2f^2)}.\nonumber
\end{eqnarray}
From this expression also one can verify that the trends are similar as above.
\subsection{Uhlmann Fidelity}
\begin{figure}[h]
\centering
\includegraphics[scale=0.65]{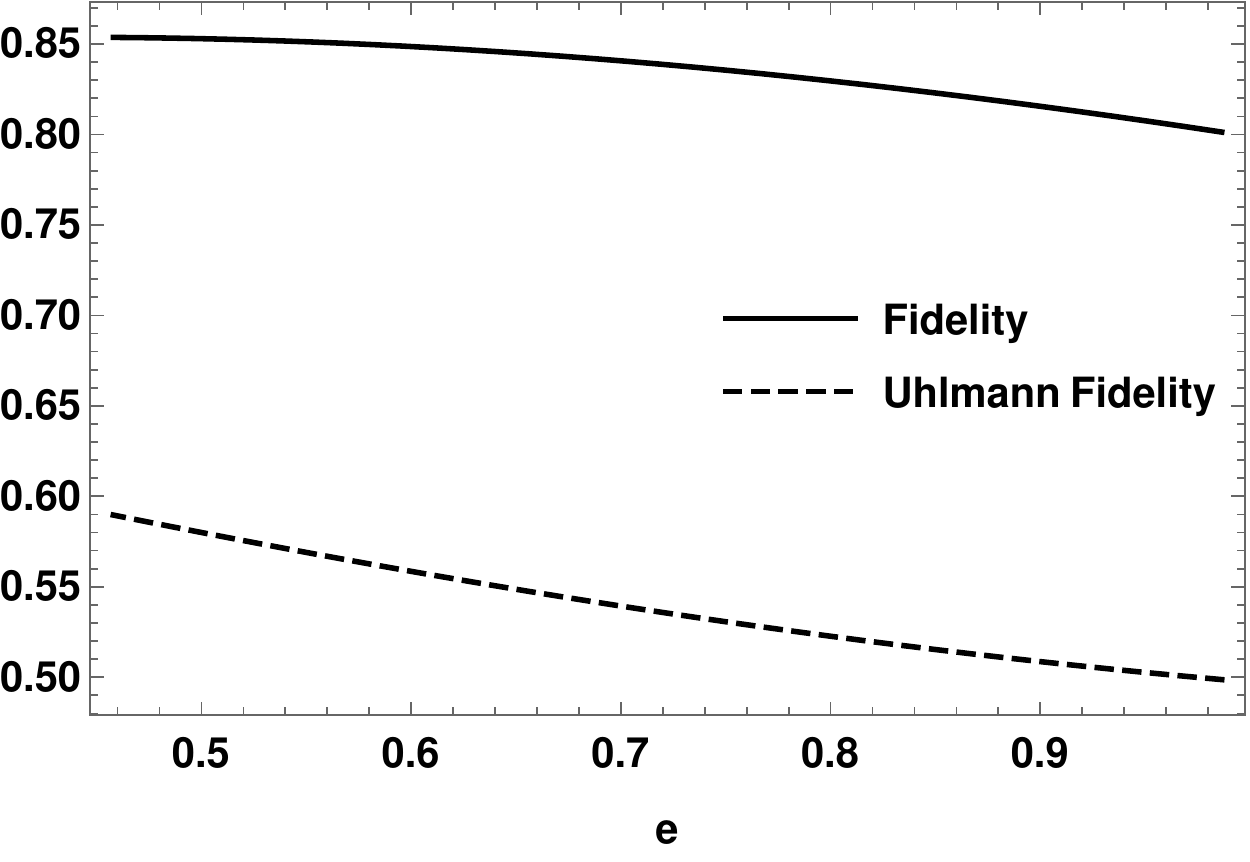}
\caption{Variation of optimal teleportation fidelity and Uhlmann fidelity with $e$ for $f=\frac{1}{2\sqrt{2}}$, $y=0$, $\mathcal{C}=0.2$ and $\mathcal{P}=0.7$ of rank-4 X-state.}
\label{fig_uhlman_fidelity}
\end{figure}

As we have seen, the optimal teleportation fidelity changes monotonically with parameters, or functions of parameters, of states. Question is apart from purity and concurrence what other physical quantities these functions of
parameters may be related to. In this subsection, we consider one such physical quantity -- Uhlmann Fidelity.  It
is known that the closeness of two states can be characterized by Uhlmann fidelity \cite{Zyczkowski2}. For two arbitrary quantum states $\rho$ and $\sigma$, the Uhlmann fidelity is defined as \cite{Zyczkowski2}
\begin{equation}\label{Ulhmann fidelity}
\mathcal{R}=\Big[{\mbox{Tr}\sqrt{{\sqrt\rho}\sigma{\sqrt\rho}}}
\Big]^2.
\end{equation}
It is a relevant quantity that describes how far apart two states are. Here we will compute the Uhlmann fidelity of a class of X-states with Bell states. As there are four Bell states, we take the maximum of the values. So we choose $\sigma$ of Eq.(\ref{Ulhmann fidelity}) as density matrices of Bell states. Then Uhlmann fidelity of the rank four X-states as in Eq. (\ref{eq:rho-Xm}) is
\begin{eqnarray}
\mathcal{R}=\max\big[&&\frac{1}{4}(1+e)\sin^2{\theta}+\sqrt{y}\cos\nu,\nonumber\\&&\frac{1}{4}(1+e)\sin^2{\theta}-\sqrt{y}\cos\nu,\nonumber\\&& \frac{1}{4}(2-(1+e)sin^2{\theta})+\sqrt{x}\cos\mu,\nonumber\\
&&\frac{1}{4}(2-(1+e)sin^2{\theta})-\sqrt{x}\cos\mu\big].\nonumber\\
&&\label{1Ulhmann fidelity}
\end{eqnarray}
There are four Uhlmann fidelities, one for each of the Bell states. Uhlmann fidelity also corresponds to the transition probability of one state to another state. We take the maximum among four, because the maximum is the most probable state. As fidelity is independent of $\mu$ and $\nu$, so without loosing generality we can choose $\mu=\nu=0$. We find out $\mathcal{R}$ explicitly with $\mathcal{C}=0.2$, $\mathcal{P}=0.7$, $f=\frac{1}{2\sqrt{2}}$ and $y=0$. For this state one can check that the Uhlmann fidelity is 
\begin{equation}\label{2Ulhmann fidelity}
\mathcal{R}=\frac{1}{4}(2-(1+e)sin^2{\theta})+\sqrt{x}.
\end{equation}
It is quite obvious from this expression that $\mathcal{R}$ is monotonically decreasing function of $e$. We have also seen
in Eq.(\ref{1fidelity rank 4}) that optimal teleportation fidelity also monotonically decreases with $e$. To visualize it we 
have plotted $\mathcal{R}$ and $\mathcal{F}$ as a function of $e$ in Fig. \ref{fig_uhlman_fidelity}.
One interpretation of the Fig. \ref{fig_uhlman_fidelity} is that the increment of parameter $e$ is somehow introducing classicality in the system. With increment of $e$, Uhlmann fidelity is decreasing i.e., the state is going far away from the Bell states and as a result optimal fidelity is also decreasing. To emphasize the importance of this quantity, we consider the states with same values of concurrence and purity that have different optimal fidelity. Let's see it
explicitly with two following states:
\begin{eqnarray}\label{examples}
&&\gamma_1=\frac{1}{3}|\psi^+\rangle\langle\psi^+|+\frac{2}{3}|\psi^-\rangle\langle\psi^-|\quad \mbox{and}\nonumber\\
&&\gamma_2=\frac{1}{3}|\psi^+\rangle\langle\psi^+|+\frac{2}{3}|00\rangle\langle00|,
\end{eqnarray}
where $\ket{\psi^\pm}=\frac{1}{\sqrt{2}}\big(\ket{01}\pm\ket{10}\big)$. These states have same purity $\frac{5}{9}$ and  concurrence $\frac{1}{3}$, but optimal fidelity is different, $\mathcal{F}(\gamma_1)=\frac{7}{9}$  and $\mathcal{F}(\gamma_2)=\frac{2}{3}$ respectively. Uhlmann fidelity of these states are given by $\frac{2}{3}$ and $\frac{1}{3}$ respectively i.e, higher Uhlmann fidelity corresponds to larger teleportation fidelity also. It seems plausible as Uhlmann fidelity is associated with distance between two density matrices. So larger distance of a state from maximally entangled state implies larger deviation from nonlocality
which in turn degrades its optimal teleportation fidelity. \\
\section{Discussion and Conclusions}\label{conclude}
We have studied nonlocal properties of  two-qubit mixed states using teleportation protocol. We 
have used a class of states, X-states. The motivation to consider X-states 
originates from the fact that for every two-qubit mixed state, there is a X-state with same 
purity and concurrence \cite{fid2}. The original state and the corresponding X-state are related 
by some global unitaries.
We have obtained the dependence of optimal teleportation fidelity on the functions of the state
parameters. All the relations for optimal fidelity indicate that noise can reduce the effectiveness of state
as a teleportation channel. In concurrence-purity region, we can find some entangled states 
for each rank which have optimal fidelity less 
than $\frac{2}{3}$. Below a certain value of purity, optimal fidelity does not increase if only concurrence is increased.
Also concurrence can not be changed arbitrarily keeping purity fixed. Moreover the amount of variation 
of optimal fidelity with purity, for fixed amount of concurrence, depends on the 
rank of the states. Our result also agrees with the work in reference \cite{Paulson}. 
Higher rank X-states give larger optimal fidelity for a fixed value of purity and concurrence.
Our investigations suggest that the nonlocal
character of a two-qubit mixed state is more involved, and require several quantities to fully characterize it.
For example, for a rank-3 X-state, the optimal fidelity depends not only on purity and concurrence,
but also on the functions $e$ and $f$. All of these quantities are functions of state parameters.
By choosing a specific set of values for the functions $e$ and $f$, we find that
the optimal fidelity changes monotonically with concurrence and purity. However, the optimal fidelity also changes monotonically with functions $e$ and $f$. This has been illustrated
in a number of plots. Thus,
these quantities also characterize the nonlocal (classical or quantum) properties of the mixed states. 
Purity and concurrence are not enough. They may characterize some average nonlocal properties. 
%We demonstrate this by obtaining functional relationship  between purity, concurrence, and a
%set of other functions a X-state parameters with the teleportation fidelity. 
%We have shown that apart from entanglement and purity of the state, other 
%functions of parameters, $e$ and $f$ also play important roles in  determining the quality of 
%teleportation. 
At some level, it is not surprising. Unlike a two-qubit pure state, a two-qubit
mixed state can have several independent parameters. However, we have found some specific
functions of the state parameters, which in addition to concurrence, also determine the nonlocal
properties of the state. The optimal teleportation fidelity changes monotonically with respect to
these functions. Interestingly, Bell violation by a X-state varies in the same way
with respect to these functions $e$ and $f$ \cite{ToBePublished}.
 These extra functions of parameters, should be related with other properties of the states which are not captured by purity, or concurrence. We have considered one such quantity, Uhlmann fidelity, and shown its importance.
There should be many more such quantities which are still to be found to fully characterize the nonlocal properties
of two-qubit mixed states.
\section*{Author contribution statement}
Development and design of the problem has been done by all the authors. S. Nandi did most of the calculations in the manuscript. All the authors contributed to the preparation of the manuscript.


\begin{thebibliography}{99}
\bibitem{fid1} C. H. Bennett, G. Brassard, C. Cr\'{e}peau, R. Jozsa, A. Peres, and W. K. Wootters, {\it Phys. Rev. Lett.} {\bf 70}, 1895 (1993).

\bibitem{ekert} A. K. Ekert, {\it Phys. Rev. Lett.} {\bf 67}, 661 (1991).

\bibitem{hillery} M. Hillery, V. Bu\u{z}ek, and A. Berthiaume, {\it Phys. Rev. A} {\bf 59}, 1829 (1999).

\bibitem{Popescu} S. Popescu, {\it Phys. Rev. Lett.} {\bf 72}, 797 (1994).

%\bibitem{Werner} R. F. Werner, {\it Phys. Rev. A} {\bf 40}, 4277 (1989).

\bibitem{Zyczkowski} K. \.{Z}yczkowski, P. Horodecki, A. Sanpera, and M. Lewenstein, {\it Phys. Rev. A} {\bf 58}, 883 (1988).

\bibitem{Bose} S. Bose, and V. Vedral, {\it Phys. Rev. A} {\bf 61}, 040101(R) (2000).

\bibitem{Paulson} K. G. Paulson, and S. V. M Satyanarayana, {\it Phys. Lett. A} {\bf 381}, 1134 (2017).

\bibitem{Verstraete} F. Verstraete, and H. Verschelde, {\it Phys. Rev. A} {\bf 66}, 022307 (2002).

\bibitem{Wooters} W. Wootters, {\it Phys. Rev. Lett.} {\bf 80}, 2245 (1998).

\bibitem{Vidal} G. Vidal, and R. F. Werner, {\it Phys. Rev. A} {\bf 65}, 032314 (2002).

%\bibitem{Horodecki} R. Horodecki, and M. Horodecki, {\it Phys. Rev. A} {\bf 54}, 1838 (1996); R. Horodecki, and P. Horodecki, {\it Phys. Lett. A} {\bf 210}, 227 (1996).

%\bibitem{spectral} In the spectral decomposition we can write a state $\rho$ as
%$\rho=\sum_{i=1}^4p_i\ket{\psi_i}\bra{\psi_i}$, with $\sum_{i=1}^4p_i=1$. So, there is total three classical mixing parameters. Each pure state contains eight parameters. Due to the overall phase factor and normalization constraint we cam reduce it to six independent parameter. Therefore, total number of parameters in $\rho$ is twenty seven. Here, all the $\ket{\psi_i}$'s are orthogonal to each other. Hence, we have total twelve orthogonality constraint. That will further reduce the parameter space of $\rho$ to fifteen. In the reference "\textit{N. Linden, S. Popescu, Fortsch. Phys. 46 (1998) 567}", authors had shown that the proper group to describe the in-equivalent states w.r.t the nonlocal properties is $U(1)\times SU(2)^n$, where $n$ is the number of parties. This group has the dimension of $3n+1$. But when we will be considering the mixed states, the phase factor (parameter for the group $U(1)$) will be canceled and we will have $3n$ parameters which can be reduced to describe the in-equivalent states. Hence, for two parties, it will reduce the parameters by $3\times 2=6$. So, $\rho$ can be represented by a nine parameter family.

\bibitem{ToBePublished} C. Datta, A. Das, S. Nandi, and P. Agrawal, In preparation.

\bibitem{Torres} J. M. Torres, and J. Z. Bern\'{a}d, {\it Phys. Rev. A} {\bf 94}, 052329 (2016).

\bibitem{fid2} P. E. M. F. Mendon\c{c}a, M. A. Marchiolli, and D. Galetti, {\it Ann. Phys} {\bf 351}, 79 (2014).

\bibitem{yu} T. Yu, and J. H. Eberly, {\it Quant. Inf. Comput.} {\bf 7}, 459 (2007).

\bibitem{x lit} A. R. P. Rau, {\it J. Phys. A-Math. Gen.} {\bf 42}, 412002 (2009); N. Quesada, A. Al-Qasimi, and D. F. V. James, {\it J. Modern Opt.} {\bf 59}, 1322 (2012); S. R. Hedemann, {\it arXiv:} 1310.7038 (2013); A. C. S. Costa, R. M. Angelo, and M. W. Beims, {\it arXiv:}1311.5702 (2013); M. Ali, A. R. P. Rau, and G. Alber, {\it Phys. Rev. A} {\bf 81}, 042105 (2010).

\bibitem{ishizaka} S. Ishizaka, and  T. Hiroshima,  {\it Phys. Rev. A} {\bf 62}, 022310 (2000).

%\bibitem{mem lit} F. Verstraete, K. Audenaert, T. D. Bie and B. D. Moor {\it Phys. Rev. A} {\bf 64}, 012316 (2001); W. J. Munro,  D. F. V. James,  A. G. White and P. G. Kwiat, {\it Phys. Rev. A} {\bf 64}, 030302 (2001);  T-C. Wei, K. Nemoto, P. M. Goldbart, P. G. Kwiat, W. J. Munro and F. Verstraete, {\it Phys. Rev. A} {\bf 67}, 022110 (2003); J. Batle, M. Casas, A. Plastino, and A. R. Plastino {\it Phys. rev. A} {\bf 71}, 024301 (2005).

\bibitem{mem exp} N. A. Peters, J. B. Altepeter, D. Branning, E. R. Jeffrey, T-C.  Wei, and P. G. Kwiat {\it Phys.  Rev.  Lett.} {\bf 92}, 133601 (2004); M. Barbieri, F. D. Martini, G. D. Nepi, and P. Mataloni, {\it Phys. Rev. Lett.} {\bf 92}, 177901 (2004). 

\bibitem{horotele} R. Horodecki, M. Horodecki, and P. Horodecki, {\it Phys. Lett. A} {\bf 222}, 21 (1996).

\bibitem{Zyczkowski2} A. Uhlmann, {\it Reports on Mathematical Physics}, {\bf 9}, 273 (1976); R. Jozsa, {\it Journal of Modern Optics}, {\bf 41}, 2315 (1994); I. Bengtsson, and K. \.{Z}yczkowski, Geometry of quantum states, Cambridge University Press, U.K, 2006.

\end{thebibliography}
\end{document}